\documentclass[twocolumn,showpacs,preprintnumbers,amsmath,amssymb,prl]{revtex4}

\usepackage{graphics}
\usepackage{epsfig}
\usepackage{bm}
\usepackage{times}

\def\beq{\begin{equation}}
\def\eeq{\end{equation}}

\def\reff#1{(\ref{#1})}

\def\subsc#1{{\mbox{\rm\scriptsize #1}}}

\def\Wcmcm{\mbox{\rm Wcm$^{-2}$}}

\def\Dhat{D_0}

\def\N3d{N_\subsc{3D}}

\def\vekt#1{\bm{#1}}

\def\vektx{\vekt{x}}

\def\vektk{\vekt{k}}

\def\vektE{\vekt{E}}

\def\vektB{\vekt{B}}

\def\vektj{\vekt{j}}

\def\halb{\frac{1}{2}}

\def\energy{{\cal{E}}}

\def\diff{\,\mbox{\rm d}}

\def\abl#1#2{\frac{\diff #1}{\diff #2}}
\def\pabl#1#2{\frac{\partial #1}{\partial #2}}

\def\imagi{\mbox{\rm i}}

\begin{document}

\title{Collisionless Absorption of Intense Laser Beams by Anharmonic Resonance}
\date{\today}
\author{P.~Mulser}
\affiliation{Theoretical Quantum Electronics, TU Darmstadt, Hochschulstr.\ 6, 64289 Darmstadt, Germany}
\author{D.~Bauer}
\affiliation{Max-Planck-Institut f\"ur Kernphysik, Postfach 103980, 69029 Heidelberg, Germany}
\author{H.~Ruhl}
\affiliation{Institute for Theoretical Physics I, Ruhr-Universit\"at Bochum,  44797 Bochum, Germany}
\date{\today}

\begin{abstract}
Two decades after the invention of chirped pulse amplification the
physical mechanism of collisionless absorption of intense laser
radiation in overdense matter is still not sufficiently well
understood. We show that anharmonic resonance in the self-generated
plasma potential of the single plasma layers (cold plasma model) or
of the individual electrons (warm plasma), respectively,
constitutes the leading physical mechanism of collisionless
absorption in an overdense plasma. Analogously to collisions,
resonance provides for the finite phase shift of the free electron
current relative to the driving laser field which is compulsory for
energy transfer from the laser beam to any medium. An efficient new
scenario of wave breaking is also indicated.
\end{abstract}


\maketitle

Collisionless absorption of intense sub-ps laser pulses in solid
targets up to 70\% and beyond is possible \cite{sauer}. The results
found their confirmation in numerous particle-in-cell (PIC)
\cite{gibbon,wilks} and Vlasov  simulations \cite{ruhlI,ruhlII}.
Since the beginning intense search for physical mechanism supporting
such high absorption values in the absence of collisions started and
a whole variety of effects was made responsible for the
observations. Pioneering work in this context has been done by
several researchers. W. Kruer and K. Eastabrook introduced the
concept of $\vektj \times\vektB$-heating in 1985 \cite{kruer} . They
realized for the first time the relevance of electron motion in
laser beam direction for heating. A further, very significant step
forward towards understanding strong collisionless absorption was
made in 1987 due to F. Brunel by recognizing the role of the
collective self-generated electric field in the plasma under oblique
incidence or the action of the $\vektj \times\vektB$-forrce,
respectively \cite{brunel}. Finally, P. Gibbon discovered that a
certain fraction of electrons pushed out into the vacuum do not
return to the target with the rhythm of the laser oscillations. For
this phenomenon, because of indicating irreversibility aspects, he
coined the term "vacuum heating" in 1994 which subsequently has
become very popular \cite{gibbon, gibbonII}. In our opinion it was
unfortunate that this phenomenon did not find the systematic
attention of researchers it deserved. So far also quantitative
analytic absorption models have been elaborated, like anomalous skin
layer absorption \cite{yang}, stochastic heating \cite{sentoku}, or
Landau damping \cite{zaretsky}, which however yield absorption
values approximately an order of magnitude lower than the values
above we are interested in in this context here. For a more detailed
analysis see \cite{bauermulser}.  According to some researchers'
view one of the phenomena makes the contribution to collisionless
absorption, according to others althogether contribute. In summary,
collisionless absorption in overdense matter is well confirmed by
experiments and simulations but not understood physically to the
desired extent.

We introduce the new concept of anharmonic resonance and show that
it is the leading physical mechanism responsible for collisionless
absorption. In a first stage we give a description of the idea for
the hurried reader; without loss of physical insight he may skip all
formulas. Then we show in detail how anharmonic resonance works with
a single plasma layer. Subsequently we study numerically the
interaction of multiple layers coupled together and, with the aid of
test particle simulations, we present the proof of our assertion for
realistic interaction, with the limitation to
 plane geometry for
being basic in many respects.

Formally, the search for collisionless absorption reduces to the
question where the phase shift $\phi$ between the monochromatic
electric field of the laser ${\vektE \sim\cos\omega t}$  of
frequency $\omega$ and the electron current density ${\vektj
\sim\sin(\omega t + \phi)}$  in Poynting's theorem comes from in the
absence of collisions since, with $\phi = 0$,
$\overline{\vektj\cdot\vektE}\sim \overline{\sin\omega t \cos\omega
t}  = 0 $. The most efficient collisionless mechanism producing
$\phi \not= 0$ is resonance. However, resonance absorption at
$\omega_p \gg \omega$ in the
 plasma has been categorically excluded so far by the
scientific community. The insight that this statement is incorrect,
because limited to the harmonic oscillator only, is the key to the
solution. In fact, when the electrons of a plane plasma layer of
thickness $d$ are displaced by an amount  $\xi = d$ or larger their
oscillation eigenperiod $T_0$ lengthens from $2\pi/\omega_p$ to
larger values and at a sufficiently big excursion $\xi$ becomes
equal to the laser period $T = 2\pi/\omega$, i.e., it enters into
resonance. This happens already at low laser intensities. The
transition from non-resonant to resonant state is irreversible and
is accompanied by the desired well-known phase shift of ${\vert
\phi\vert= \pi}$. This behavior persists also in the real plasma
with crossing of multiple oscillators (breaking) and mutually
influencing each other thereby lowering the single layer resonance
threshold. Owing to the non-integrable (chaotic) nature of the
governing equations already in the simplest case this final part of
the assertion can be proved only with the help of numerical methods.

\begin{figure}
\includegraphics[width=0.475\textwidth]{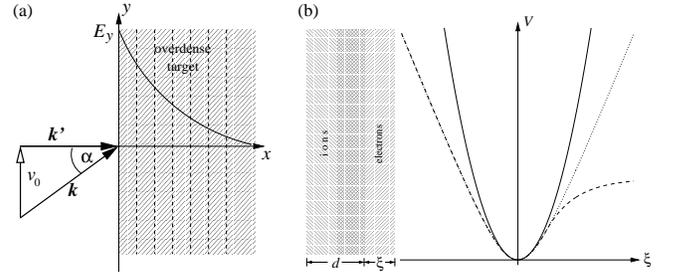}
\caption{(a) Oblique incidence of parallel laser beam of wave vector
$\vektk$ in the lab frame and $\vektk'\parallel x$ in the system
boosted by $v_0 = c\sin\alpha$. The overdense target is cut into
layers of thickness $d$. (b) Large electron displacement $\xi$ in
plasma layer (LHS) and anharmonic closed (dotted) and semi-open
(dashed) potentials. In wider potential than harmonic (solid) the
frequency decreases with increasing $\xi$. \label{fig1}}
\end{figure}

Now we proceed to the quantitative proof. A plane, fully ionized
target is assumed to fill the half space $x > 0$. A plane wave
$\vektE(\vektx,t)  = \vektE_0 \exp[\imagi(\vektk\cdot\vektx - \omega
t)]$ in p-polarization in $y$-direction, wave vector $\vektk$ and
frequency $\omega$, is incident from $-\infty$ under an angle
$\alpha$ onto the plasma surface (Fig.~\ref{fig1}a). After applying
a Lorentz  boost $v_0 = c \sin\alpha$ the wave impinges normally.
The target is thought to be cut into a sufficient number of thin
parallel layers, each of which is exposed to a driving force $F$
acting along $x$ of magnitude $ev_0B$ and frequency $\omega$, and an
additional component originating from the Lorentz force of the
oscillatory motion along $y$ of frequency $2\omega$ ($e$ electron
charge, $B$ magnetic field of the laser). With the electron
displacement $\xi$ in $x$-direction and immobile ions of density
$n_0$ (corresponding to $\omega_{p0}$)  the motion of a single layer
(Fig.~\ref{fig1}b) in the non-relativistic limit is well determined
by the potential $V = m_e(\omega_{p0} d/2)^2 (1 + \zeta^2)^{1/2}$,
$\zeta = 2\xi/d$, $m_e$ electron mass (derivation is elementary).
Hence,  the plasma oscillator is governed by the dimensionless
equation \beq \abl{^2\zeta}{\tau^2} + \pabl{(1 +
\zeta^2)^{1/2}}{\zeta} = D(\tau) \label{eom}\eeq where $\tau =
\omega_{p0}t$ and $D=2F/(\omega_{p0}^2 d)$.  The eigenperiod \beq
T_0 = \int 4[2(\energy - V(\xi))/m_e]^{-1/2}\,\diff \xi \eeq for a
fixed energy $\energy$, at high excitation and $\omega_{p0}\gg
\omega$ is given by $T_0 = 8(\omega_{p0}^2d)^{-1/2} \xi_0^{1/2}$.
The eigenfrequency  $\omega_0 = 2\pi/T_0  = (\pi/4)
(\omega_{p0}^2d)^{1/2}  \xi_0^{-1/2}$  decreases with increasing
oscillation amplitude $\xi_0$, in contrast to the linear harmonic
oscillator. To make the essence clear we concentrate on large angles
$\alpha$ and set $D = a_0(t)\cos\omega t$, $a_0(t)$ amplitude. At
$\omega_0= \omega$ the single plasma layer is resonantly excited at
amplitude $\xi_0 = \xi_r = (\pi\omega_{p0}/4\omega)^2d$ and $t =
t_r$.  Under the assumption that the oscillator is driven into
resonance during half a laser cycle, i.e., when the driver amplitude
$E_0 = m_e \omega_{p0}^2d/(4e)$, with $d = 0.1$--$0.2$~nm and
$\omega_{p0} = 2\times 10^{16}$\,s$^{-1}$ this happens at the  laser
intensity  $I = 10^{15}$\,\Wcmcm.

\begin{figure}
\includegraphics[width=0.45\textwidth]{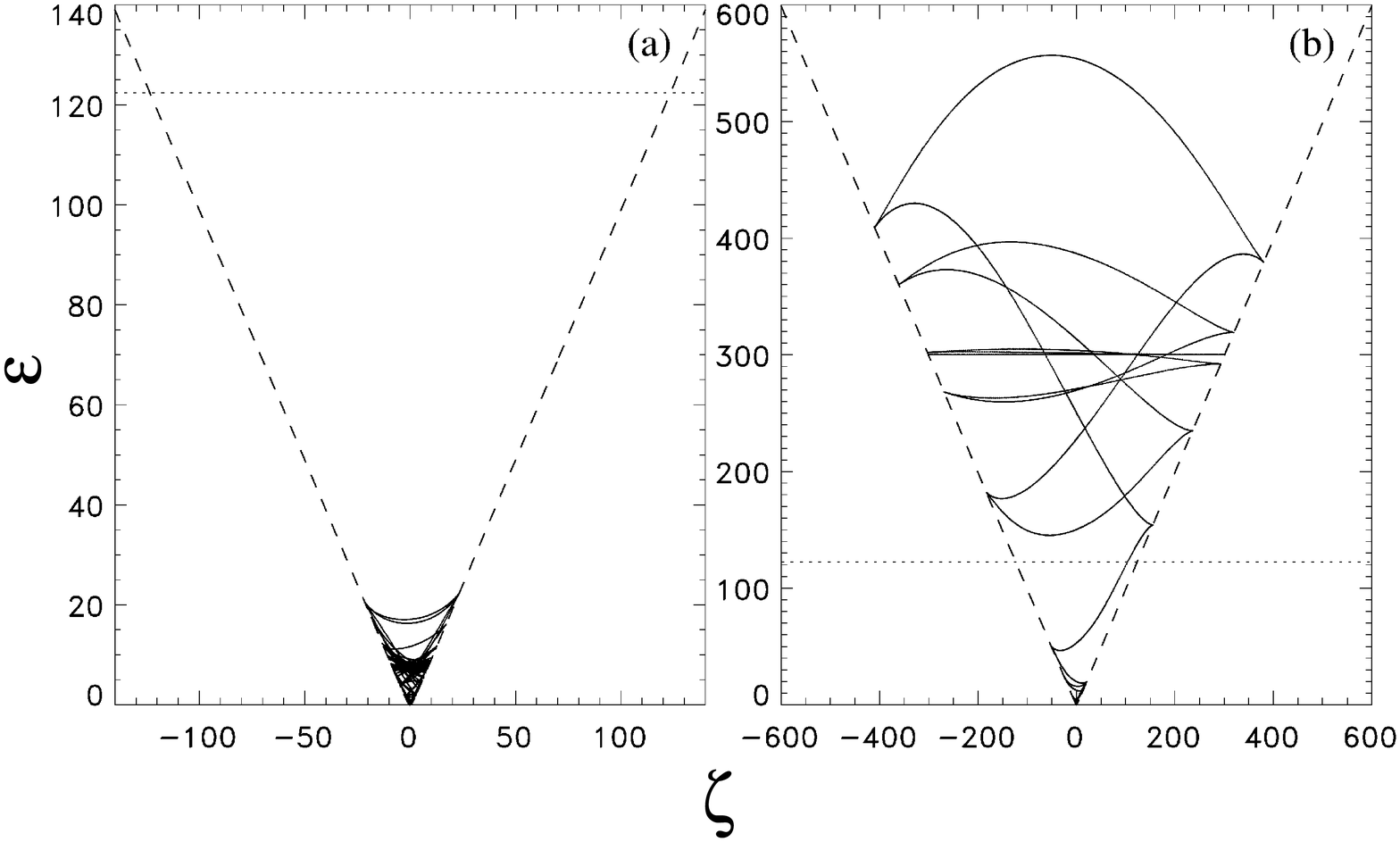}
\caption{Excitation $\varepsilon(\zeta)=\dot{\zeta}^2/2+V(\zeta)$ of
the oscillator from Eq.~\reff{eom} by the driver
$D(\tau)=\Dhat\sin^2[\omega \tau/(2N\omega_{p0})]\cos\omega
\tau/\omega_{p0}$, $\omega/\omega_{p0} = 0.1$, $N=20$, (a) below
resonance ($\Dhat=0.921$), (b) above resonance ($\Dhat=0.923$) with
an $43$ times higher final energy gain $\varepsilon_f$ than in (a)
although the driver strength is changed only by $0.2$\%. The
potential and the resonant energy level  are indicated dashed and
dotted, respectively. Note the different scales in a,
b.\label{fig3}}
\end{figure}

Under a weak driver $D(\tau)$ in ~\reff{eom} the eigenfrequency
$\omega_0$ is much higher than $\omega$  and the excursion
$\zeta(\tau) \sim \cos(\omega t + \varphi)$ follows adiabatically
$D(\tau)$, i.e., $\varphi=0$. In other words, $\zeta(\tau)$ moves in
phase with $D(\tau)$. At large displacements the restoring force
weakens and $\omega_0 \ll \omega$, thus imposing $\vert\varphi\vert
= \pi$, like a free electron oscillating in the laser field. At an
intermediate $\zeta$-value the phase shift must have taken place
with the consequence that the product $\dot{\zeta}D $ transits from
$\sim\sin\tau \times \cos\tau$ to $~\cos^2\tau$. It is the resonance
region. Only there, in the so-called neighborhood of stationary
phase \cite{mathews} resonance occurs and irreversible energy
absorption by the oscillator is possible. A necessary condition for
stationary phase is the existence of a point with $\vert\varphi\vert
= \pi/2$. Slow transitions through resonance can be visualized
geometrically with the help of the Cornu spiral as explicitly shown
for $\omega_0(t)$ in another context in \cite{mulser}. In order to
get rid of such WKB-like excitation Eq.~\reff{eom} is solved
numerically with a smooth $N$-cycle $\sin^2$-shaped driver $D=a_0(t)
\cos\omega t =\Dhat\sin^2[\omega t/(2N)]\cos\omega t$. In
Fig.~\ref{fig3}a with a driver $\Dhat = 0.921$ the layer remains
below resonance, the energy gain
$\varepsilon_f=\varepsilon(2N\pi\omega_{p0}/\omega)$ after the pulse
is negligible. By an increase of the driver of only $\Delta D_0 =
0.002$ resonance takes place. Due to dephasing above resonance the
energy content shows the typical periodic variations attenuating in
time towards a finite value when the pulse is over. Much energy is
stored now in the oscillator (see the horizontal orbits in
Fig.~\ref{fig3}b). Compared to case (a) the energy gain is increased
by a factor of $43$.

\begin{figure}
\includegraphics[width=0.4\textwidth]{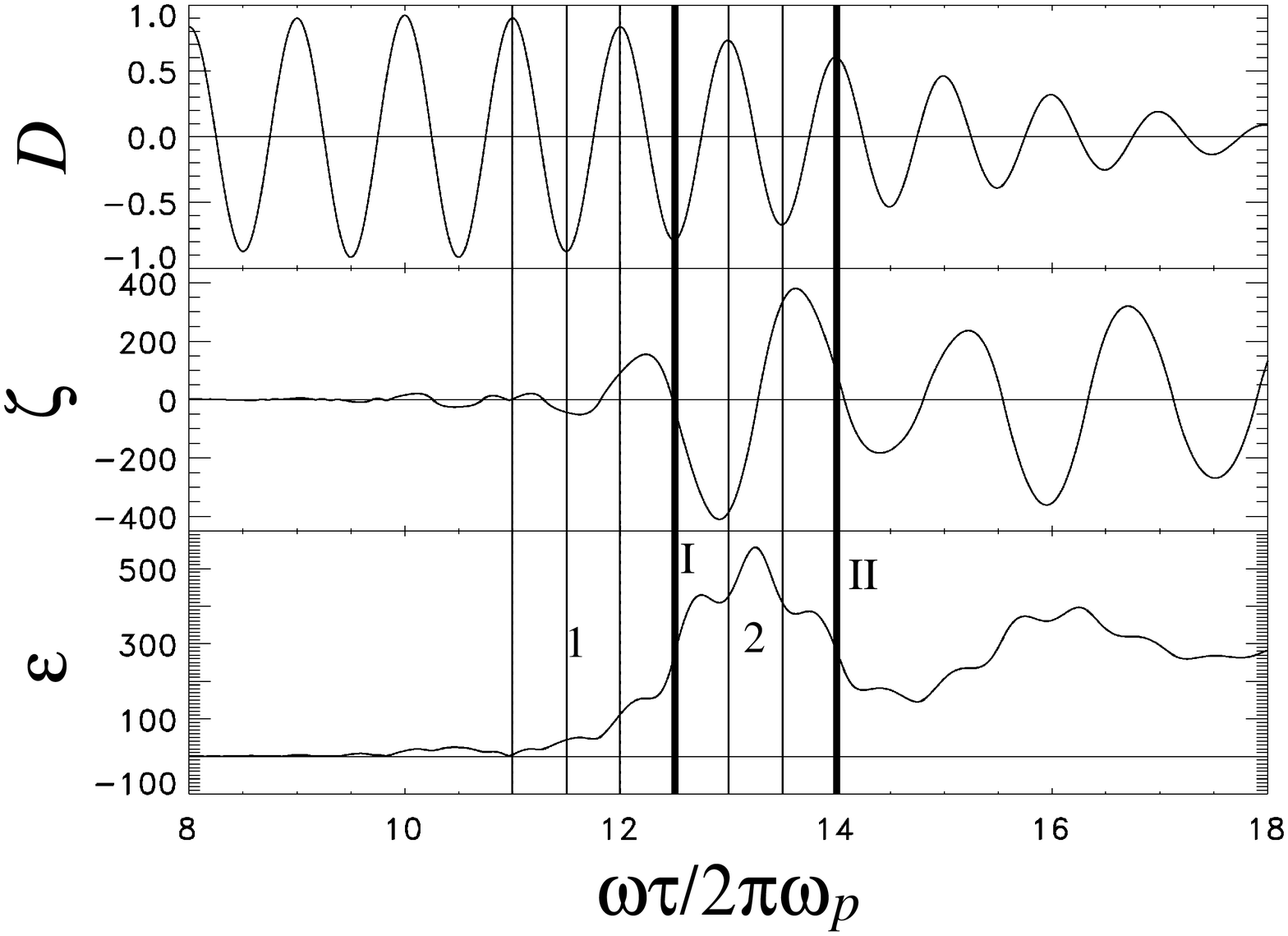}
\caption{Driver $D(\tau)$, excursion $\zeta(\tau)$ and absorbed
energy $\varepsilon(\tau)$ vs time (in driver cycles) for the case
in Fig.~\ref{fig3}. Each time resonance is crossed $\zeta$ undergoes
a phase shift by $\pi$. Bold lines I, II indicate instants of phase
lag $\zeta$ are $\pm\pi/2$ between $D$ and $\zeta$ , i.e., maximum
energy gain and loss (points of stationary phase); modulations in
$\varepsilon$ originate from the $\omega + \omega_0$ spectral
component. \label{fig4}}
\end{figure}

Figure~\ref{fig4} is of particular relevance. It shows the driving
field $D(\tau)$, the displacement $\zeta(\tau)$, and the energy
gained $\varepsilon(\tau)$. At position $1$ the oscillator is
entering resonance ($\varepsilon$ starts increasing, $\omega_0 >
\omega$), $D$ and $\zeta$ are in phase; at $2$ it is leaving
resonance ($\varepsilon$ starts decreasing, $\omega_0 < \omega$),
$D$ and $\zeta$ are dephased by $\pi$. Positions I and II (points of
stationary phase) indicate maximum energy gain and maximum energy
loss, $D$ and $\zeta$ are dephased by $\pm{\pi/2}$. Thus, the
resonance signature is preserved in a rapid transition. The
phenomenon repeats in the second maximum of $\varepsilon$, etc. The
definition of resonance as the points of stationary phases is the
natural extension of the concept of resonance to anharmonic and
nonlinear oscillators. An equivalent criterion to be used later is
this: (i) the half widths of the local maxima of $\varepsilon(\tau)$
is not much shorter than half a driver cycle $T = 2\pi/\omega$ (in
Fig.~\ref{fig3} it is $1.5T$), in contrast to a collisional event
which is almost instantaneous, and (ii) $\varepsilon (\tau
\rightarrow \infty)
> 0$.

\begin{figure}
\includegraphics[width=0.45\textwidth]{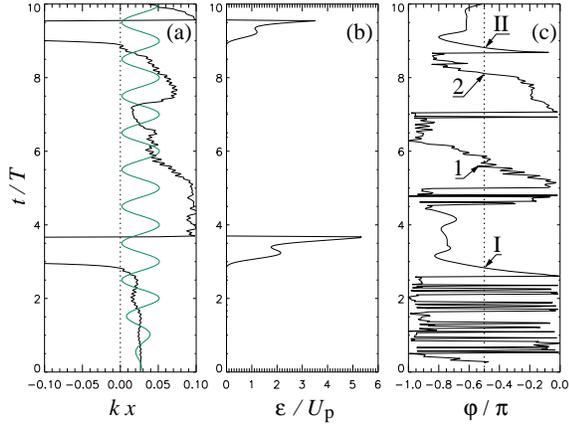}
\caption{Resonance dynamics of layer \#32 as a function of laser
cycles. Position $kx$ (a), $k$ wave number, absorbed energy
$\varepsilon$ (b) and phase $\varphi$ between velocity $v$ and
driving laser field $E = E_0(x)\cos\omega t$ (mapped into the
interval $[-\pi,0]$) (c); $\varepsilon$ in units of ponderomotive
potential $U_p$. Passages through $-\pi/2$ are indicated by I, II,
and 1 and 2 (see text). $E_0(x)$ is calculated selfconsistently and
corresponds to laser intensity $I = 3.5\times 10^{18}$\,\Wcmcm.
After resonance at $3.5$ laser cycles the layer is pushed back into
the target with high velocity ("disruption"). After crossing the
opposite target surface the layer is substituted by a new layer.
Note $E \sim -D$. \label{fig5}}
\end{figure}

The motion of $N$ layers of free electrons and ions is described for
arbitrarily large oscillations by the non-separable (non-integrable,
chaotic), yet elementary Hamiltonian
 \beq H=\sum_{k=1}^N \left(\frac{p_k^2}{2} + \halb \sum_{k'\neq
k}^N V_{kk'} + \sum_{l=1}^N V_{kl} - D(\tau) \zeta_k \right) \eeq
with $p_k=\diff{\zeta}_k/\diff\tau$,
$V_{kk'}=[1+(\zeta_k-\zeta_{k'})^2]^{1/2}$,
$V_{kl}=-[1+(\zeta_k-\zeta_{0l})^2]^{1/2}$, $\zeta_k=2x_k/d$,
$\zeta_{0l}=2a_l/d$. When one of the layers is driven into resonance
it starts moving opposite to the coherently moving non-resonant
layers thereby crossing one or several adjacent oscillators. This is
a new scenario of very effective wave breaking not described in the
literature so far. We give it the name of resonant (wave) breaking.
It means loss of coherence and leads to flattening of the collective
potential and, in concomitance, to a reduction of the resonance
threshold. As a representative case we study the dynamics of a 100
times overdense target, subdivided into 120 layers of $ d = 0.125 $
nm each, on which $I = 3.5\times 10^{18}$ \,\Wcmcm at $\lambda =
800$ nm is impinging. The typical scenario is as follows: After
being pulled out into the vacuum and oscillating there for some time
the layers are pushed back in a disruption-like manner into the
target (formation of jets). When the layers leave from the back of
the target they are replaced by new layers with zero momentum (cold
return current). First indication of resonance: After 100 laser
cycles more than half of the now more than 2700 layers have gained
energies exceeding their quiver energy $U_p$. The power spectrum
shows a plateau between $1 U_p$ and the cut off at $6 U_p$. To show
the occurrence of resonance explicitly the phase of each layer with
respect  to the driving laser field was investigated. A typical
example is shown in Fig.~\ref{fig5} with layer \#32, LHS trajectory
and driver field (a), middle absorbed energy (b), RHS phase
$\varphi$ of velocity $v \sim \sin(\omega t + \varphi)$ with respect
to the driver $E \sim \cos\omega t$ (c). Over $T/2$ there is a
continuous and smooth transition of $\varphi$ through $-\pi/2$ at
$t/T = 2.8$ (I) with a simultaneous strong increase in the absorbed
energy (b) and the excursion $x$, with following disruption of the
layer at $t/T = 3.6$. The change of $\varphi$ is clearly seen also
in (a). Another resonance of the same kind is found  at $t/T = 8.8$
(II). Other two passages of $\varphi$ through $-\pi/2$ at $t/T =
5.7$ (1) and $8.1$ (2) show rapid fluctuations and hence do show
almost no energy gain and no disruption [see (b), (a)]. Transitions
of this latter kind are morphologically clearly distinguishable from
the former case, and for none of the 120 layers they are able to
accelerate them across the target. This proofs in an impressive way
that our definition of resonance as points of stationary phase and
the resonance criterion (i)+(ii) are correct and meaningful.  We
conclude that in the cold plasma model all absorption is by
anharmonic resonance. The layers disrupt in the chaotic order 6, 5,
4, 3, 2, 15, 17, 16, 9, 11, 10, 8, 12, 24, 31, 34, 28, 32, 14, 30,
1, 33, 45, 26, 36, 27, etc.; layer 113 disrupts before front layer
0. This is one of the fundamental differences in the dynamics in
comparison to [16]. More essential is that the Hamiltonian Eq. (3)
acts on the single layer like a half open potential, as sketched by
the dashed line in Fig.~ \ref{fig1}b. We tested this explicitly and
found that the degree of absorption does not depend much on the
hight of the potential barrier, even when this latter lies
considerably below the resonance level of the closed potential. Only
when it is set equal to zero as in \cite{brunel} the absorption
almost vanishes because of the absence of resonance. The half open
potential with finite threshold towards the target interior is
clearly seen in our PIC simulations with the PSC code.

\begin{figure}
\includegraphics[width=0.5\textwidth]{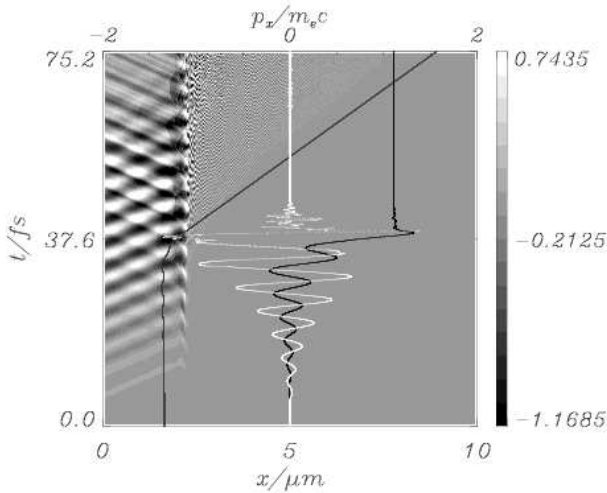}
\caption{1D3V PIC simulation of a Nd laser beam interacting with a
thick overdense target with mobile ions in the boosted frame
(parameters see text). Regular shadow structure: laser field, LHS
black trajectory: displacement $x(t)$ of test electron \#4, RHS
black trajectory starting from $0 m_e c$: corresponding momentum
$p_x$, white line: total electric field at position $x(t)$.
Resonance (strong momentum increase and phase shift $\varphi(t)$)
and disruption are impressive. No disruption without preceding
resonance ever seen in any test. All resonances look very similar to
each other. \label{fig6}}
\end{figure}

The final step of the proof , i.e., high degree of absorption as
mentioned in the introduction \cite{sauer, gibbon, wilks, ruhlI} and
absorption itself being accomplished by anharmonic resonance, at
present can only be based on computer simulations.  For this purpose
a $10^6$ particles PIC simulation with the PSC code \cite{ruhlIII}
in its collisionless mode under $45^\circ$ \ irradiation is
performed. In the boosted frame a Gaussian Nd laser beam of $I =
10^{17}$ \,\Wcmcm\ and halfwidth of 26 fs acts on a plane $80$ times
overcritical $7.3 \mu$m thick target. The orbits of 200 test
electrons equally distributed over the target are followed during 15
laser cycles of $T_{Nd} = 5$ fs. Again, the outcome of the
statistics is overwhelming: all test electrons interacting with the
laser field are resonantly accelerated and disrupt nearly
immediately. In Fig.~\ref{fig6} the time history of test electron is
depicted. The electron enters the laser field (shadowed interference
pattern) and interacts resonantly (see momentum) and escapes into
the target an instant later with $5$ times $U_p$. All resonant
orbits look nearly identical to each other. The resonance character
is ensured by the high energy gain and the phase shift in comparison
to the total laser field (white line superposed). The shadowed fine
structure right of the laser field is due to plasmons. The
simulation tells also important details on the heating mechanism.
The primary effect is the generation of the fast spectrum by
resonance. They generate "solitary", i.e. non-Bohm-Gross plasmons in
the dense interior which, in turn, heat cold electrons by a
mechanism resembling  Landau damping. This completes our last part
of the proof. It becomes clear now why perturbative theories when
starting in zero approximation  from straight orbits, e.g.,
anomalous skin effect, fail to explain strong absorption because
anharmonic resonance is outside the validity of a linearized
treatment of standard type. We want to point out that this mechanism
is active also in  long fs or ps pulses when profile steepening is
so strong that no linear resonance can take place.

In summary we have addressed the leading physical mechanism of
collisionless absorption and, in particular, we have discovered the
phase shift between driver field and electron current, indispensable
for absorption, as a resonance effect. The discovery may be viewed
as the extension of the well-known fact that a single point charge
cannot absorb a photon unless it resonates in an outer potential. As
a byproduct we have found a new, very efficient scenario leading to
(wave) breaking. Our results will have a major impact on the further
progress in the theory of laser-dense matter interaction. So, for
example, on the basis of anharmonic resonance the appearance of a
hot temperature lying higher, often considerably higher, than the
mean electron quiver energy and the fact that the Maxwellian-like
tail of their energy spectrum is filled up nearly instantaneously
(fs time scale) finds a very natural explanation. The latter
phenomenon, never decently discussed in the literature, is
particularly surprising as the fast electrons do almost not interact
together. Finally, as the overwhelming majority of oscillatory
motions in nature are anharmonic, the harmonic oscillator being the
great exception, the model developed here will find its application
in various other fields of fundamental and applied science , e.g.,
formation of cracks by fatigue under oscillatory stress, and in
catastrophe theory. In the specific field of high power - overdense
matter interaction the main relevance of our finding we see in the
possibility to tailor the electron spectrum for various applications
(electron and ion acceleration, fast ignition, etc.) by designing
targets properly, for instance by choosing carefully their
thickness. The energy spectrum of thin targets is more energetic
owing to multiple resonances than the thick target spectrum.
Finally, the simulations have also revealed that in the latter case
the spectrum is subject to continuous metamorphosis in time, an
aspect which may play an important role in fast ignition.

This work was supported by the ILIAS Programme at Gesellschaft f\"ur
Schwerionenforschung (GSI) and the Deutsche Forschungsgemeinschaft.



\end{document}